# Suppressing Material Loss for Functional Nanophotonics Using Bandgap Engineering

Mingsong Wang[1,2], Alex Krasnok[2,*], Sergey Lepeshov[3], Guangwei Hu[2,5], Taizhi Jiang[4], Jie Fang[1], Brian A. Korgel[4], Andrea Alù[2,*], and Yuebing Zheng[1*]

[1]*Walker Department of Mechanical Engineering and Texas Materials Institute, The University of Texas at Austin, Austin, Texas 78712, United States*
[2]*Photonics Initiative, Advanced Science Research Center, City University of New York, New York 10031, USA*
[3]*ITMO University, St. Petersburg 197101, Russia*
[4]*McKetta Department of Chemical Engineering, The University of Texas at Austin, Austin, Texas 78712, United States.*
[5]*Department of Electrical and Computer Engineering, National University of Singapore, Singapore 117583, Singapore.*

*akrasnok@gc.cuny.edu, aalu@gc.cuny.edu, zheng@austin.utexas.edu*

## Abstract

*All-dielectric nanoantennas have recently opened exciting opportunities for functional nanophotonics, owing to their strong optical resonances along with low material loss in the near-infrared range. Pushing these concepts to the visible range is hindered by a larger absorption coefficient of Si and other high-index semiconductors, thus encouraging the search for alternative dielectrics for nanophotonics. In this paper, we employ bandgap engineering to synthesize hydrogenated amorphous Si nanoparticles (a-Si:H NPs) offering ideal features for functional nanophotonics. We observe significant material loss suppression in a-Si:H NPs in the visible range caused by hydrogenation-induced bandgap renormalization, producing resonant modes in single a-Si:H NPs with Q factors up to ~100, in the visible and near-IR range for the first time. In order to demonstrate light-matter interaction enhancement, we realize highly tunable all-dielectric nanoantennas coupling them to photochromic spiropyran (SP) molecules. We show ~70% reversible all-optical tuning of light scattering at the high-Q resonant mode, along with minor tunability when out of resonance. This remarkable all-optical tuning effect is achieved under a low incident light intensity ~3.8 W/cm$^2$ for UV light and ~1.1×10$^2$ W/cm$^2$ for green light.*

Light trapping is crucial in a plethora of photonic applications, including lasers[1], sensing[2], and harmonic generation[3], to name just a few. High-quality resonators facilitate light trapping in the form of localized resonant modes for a time $\sim 2Q/\omega_0$, where $Q$ is the quality factor of the mode and $\omega_0$ is its eigenfrequency. In practice, the Q factor of a single standing resonance can be defined via its resonance linewidth at half maximum ($\Delta\omega$) as $Q = \omega_0/\Delta\omega$, implying that a higher Q factor possesses narrower resonant lines. Optical resonators supporting high-Q modes include microdisk resonators[4–8], microspheres[9], Bragg reflector microcavities[10], and photonic crystals[11–13], whose high Q factors can go up to $\sim 10^3 - 10^6$. However, these microscale dielectric resonators are bulky and exhibit modest light-matter interactions when averaged over their large sizes.

Functional nanophotonics requires light localization and hence high Q factors in subwavelength optical resonators. To this end, nanoscale light trapping based on plasmonic resonances of metal nanoparticles[14–21] and on Mie resonances of high-index dielectric nanostructures has been explored. However, the Q factor of these nanoresonators is limited by radiative and material losses and it does not exceed a few tens in the visible range. Although radiative losses can be suppressed with the tailoring of weakly scattering states, such as bound states in the continuum (BICs)[22–25], the material (dissipation) loss requires a fundamentally different approach. High-index dielectric resonant NPs look especially promising due to their lower material loss compared to plasmonic nanoparticles[14–19,26–29], enabling low-loss photonic devices, e.g., nanoantennas, metalenses, and lasers[30–37]. Recent studies have also revealed the importance of high-index dielectric nanoantennas as versatile platforms for active nanophotonics[38–43]. In contrast to plasmonic structures, where tailoring of the magnetic optical response is a challenging task, the response of high-index dielectric NPs is governed by both magnetic and electric resonances in visible and NIR range. Among high-index dielectrics, Si-based NPs have been of particular interest for nanophotonic applications[30] because Si is the choice material for the established semiconductor industry, where fully developed processing and characterization tools can readily be applied to create future CMOS-compatible integrated photonic and electronic systems based on a single material platform[30,44,45].

Theoretical studies have predicted enhanced Q factors for high-order multipole modes in low-loss high-index dielectric NPs compared with the fundamental dipole mode, because of reduced radiation loss[35,44,46,47]. However, such modes are fragile and very sensitive to material losses, hence they are hardly observable in practice. For example, crystalline silicon (c-Si) has a broad optical absorption in the wavelength range from 300 nm to 1150 nm, with an absorption coefficient larger than $1\times 10^3$ cm$^{-1}$ over the whole visible region[48] that reduces Q factor of the fundamental magnetic dipole (MD)[30] and suppresses all high-order modes. This obstacle has prevented the development of low-loss and compact photonic devices for the visible region and prompts a further search for alternative materials.

Recent studies have revealed the power of bandgap engineering for tailoring materials with superb optical properties[49]. For example, in Ref.[50] it has been reported that increasing the atom spacing may significantly reduce material loss in metal nanoparticles. Another example is the optical properties of phase change materials, in which the amorphous phase exhibits lower material losses[51]. In this paper, we employ bandgap engineering to tailor hydrogenated amorphous Si nanoparticles (i.e., a-Si:H NPs) and endow them with low dissipation in the visible range. First, we controllably vary the hydrogen concentration and the crystalline structure from crystal to amorphous and demonstrate that, similar to the amorphous phase of phase change materials, a-Si:H NPs exhibit smaller material loss. Next, based on single-nanoparticle spectroscopy of a-Si:H NPs, we report the first experimental observation of strong magnetic quadruple (MQ) and magnetic octupole (MO) scattering in the visible and NIR regions in subwavelength nanoparticles. Interestingly, our single-nanoparticle scattering spectra reveal that MO scattering modes have a Q factor up to 100, which, to the best of our knowledge, is the highest ever measured for dielectric NPs studied so far. Given the enhanced light-matter interactions, we are able to observe high tunability of the supported resonances by simply adjusting the hydrogen concentration. Finally, in order to demonstrate the unique opportunities offered by our approach to low-loss functional nanophotonics, we show that the coupling of such high-Q scattering resonances to a single a-Si:H nanoantenna with spiropyran (SP) photochromic molecules results in switchable ~70% tuning of the scattering peak intensity upon changing the state of the photochromic molecules between transparent SP and colored merocyanine (MC) states. This significant all-optical tunability is achieved using a non-laser light source with low incident light intensity.

**Bandgap Engineering of Low-Loss Hydrogenated a-Si Nanoparticles**

It is known that the bandgap in Si is induced from short-range ordered structures of neighboring Si atoms, which lead to $sp^3$-hybridisation of s and p valence electrons, forming two bands with decreasing interatomic distance and an indirect bandgap. The bandgap energy of a-Si:H, therefore, depends on the interaction between the $sp^3$ hybrids of neighboring Si atoms (Fig. 1a), and it is sensitive to the interatomic distance of Si-Si bonds[52,53]. As schematically shown in Fig. 1b, the voids in a-Si:H brought by hydrogen atoms cause a distortion of Si-Si bond, which leads to different bonding angles, smaller Si-Si distances and thus a larger bandgap, compared to c-Si. c-Si has an indirect bandgap ($E_g$) of 1.16 eV (which corresponds to the wavelength in free space of 1069 nm) and a forbidden direct transition ($E_g'$) of 3.2 eV (wavelength is 387 nm)[48]. As we demonstrate below, by tuning the hydrogen concentration, a-Si:H NPs can obtain the absorption bandgap larger than 1.77 eV (700 nm), and thus sustain a lower loss and strong higher-order scattering modes with high Q factors in the visible range.

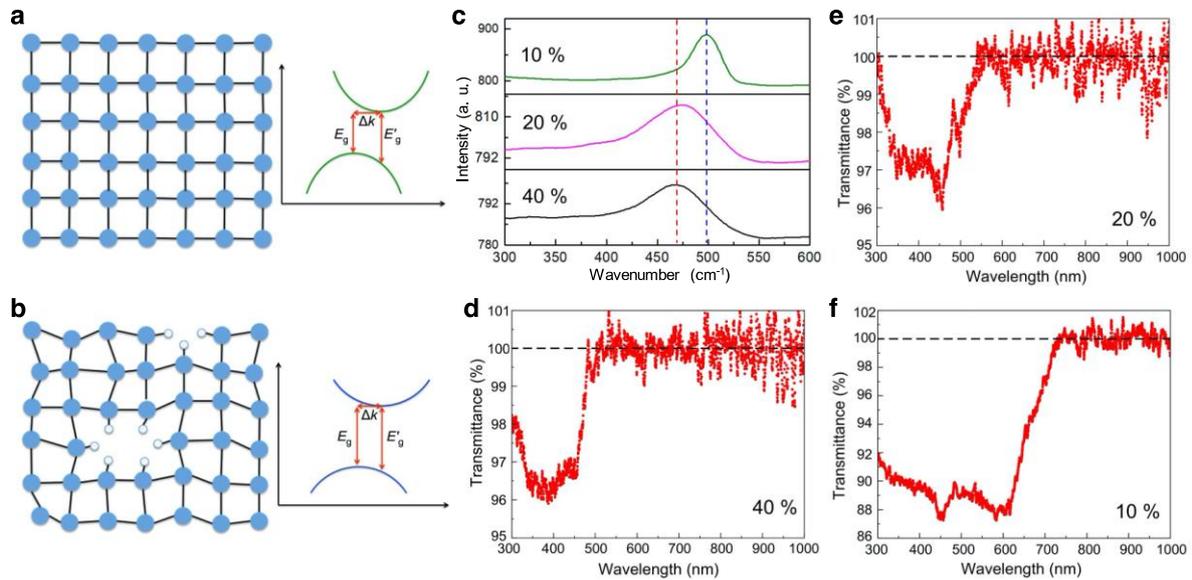

**Fig. 1 | Bandgap engineering of low-loss hydrogenated a-Si nanoparticles. a,b,** Two-dimensional schematic representations of the atomic arrangement and the bandgap of c-Si (**a**) and a-Si:H (**b**). $E_g$ is the indirect bandgap, $E_g'$ is the direct transition and $\Delta k$ is the wave-vector mismatch. **c,** Raman spectra of a-Si:H NPs with 40%, 20%, 10% of hydrogen. The blue dashed line indicates the peak position of 495 cm$^{-1}$ and the red dashed line labels the peak position of 470 cm$^{-1}$. **d,e,f,** Transmission spectra of a-Si:H NP(40) (**d**), a-Si:H NP(20) (**e**) and a-Si:H NP(10) (**f**) in ethanol after removing scattering signals.

The a-Si:H NPs studied here were fabricated by chemical synthesis (detailed information is in Methods) [52,54]. The SEM images of a-Si:H NPs are shown in Supplementary Figs. 1-3. Raman spectroscopy was employed to verify that the distortion of the Si-Si bond and the amorphous nature of a-Si:H NPs increases with an increase of hydrogen concentration. As shown in Fig. 1c, when the hydrogen concentration is 10 at.%, the Raman peak is at about 495 cm$^{-1}$ and it is relatively narrow (see the green spectrum in Fig. 1c). This peak is brought by the appearance of 1-2 nm nanocrystalline Si domains in the amorphous matrix[52]. As the hydrogen concentration increases, the Raman peak becomes broader and shifts to ~470 cm$^{-1}$, which is associated with a transverse optical phonon mode of a-Si and stretching vibrational modes of Si-Si bonds in a silicon tetrahedron, thus indicating an increase in the disordered nature of a-Si:H NPs[52].

The bandgap of a-Si:H NPs with different hydrogen concentrations is determined from measured transmission spectra of a-Si:H NPs, which are presented in Fig. 1d-f. Scattered signals are subtracted to obtain pure absorption signals (more detailed information is in Supplementary information section 2). Samples with 40 at.%, 20 at.% and 10 at.% hydrogen (a-Si:H NP(40), a-Si:H NP(20) and a-Si:H NP(10)) have absorption bands starting at about 500 nm, 560 nm, and 710 nm, respectively. This confirms that the absorption bandgap of a-Si:H broadens as the hydrogen concentration decreases, and it can be tuned by varying hydrogen concentration. It should be mentioned that the transmission spectra are obtained through ensemble measurements, so each individual a-Si:H NP may have a slightly different absorption peak than those in Fig. 1d-f.

Dark-field scattering spectra of single-NPs were measured (see Supplementary Fig. 6 for the dark-field setup). Figs. 2a-c show the scattering spectra of a single a-Si:H NP(40), a-Si:H NP(20), and a-Si:H NP(10) of diameters 370 nm, 320 nm, and 414 nm, respectively. Three major peaks in each spectrum are observed, which were fitted into three Lorentzian peaks associated, as proven by our theoretical analysis presented below, with magnetic quadrupole (MQ) resonances at 639 nm (Fig. 2a) and 717 nm (Fig. 2b) and 961 nm (Fig. 2c), electric dipole (ED) resonances at 730 nm (Fig. 2a) and 771 nm (Fig. 2b), magnetic dipole (MD) resonances at 865 (Fig. 2a) nm and 937 nm (Fig. 2b), a magnetic octupole (MO) resonance at 771 nm (Fig. 2c), and a electric quadrupole (EQ) resonance at 807 nm (Fig. 2c), respectively. Remarkably, different from any previously reported SiNPs, MQ and MO modes in our a-Si:H NPs show a large scattering cross-section (see Fig. 2b and c, blue fitting curves). The strong MQ and MO scattering peaks are repeatable and consistent with the prediction from Mie scattering theory[47], confirming the low dissipative nature of a-Si:H NPs in the visible and NIR range. It should be noted that the refractive index of a-Si:H NPs decreases with increased hydrogen concentration, due to the density reduction brought by hydrogenated nanovoids[53]. This causes a-Si:H NPs to have a smaller refractive index than a-SiNPs. The higher hydrogen concentration requires the size of a-Si:H NPs to be larger than the one of SiNPs to keep the Mie resonance peak positions aligned[35].

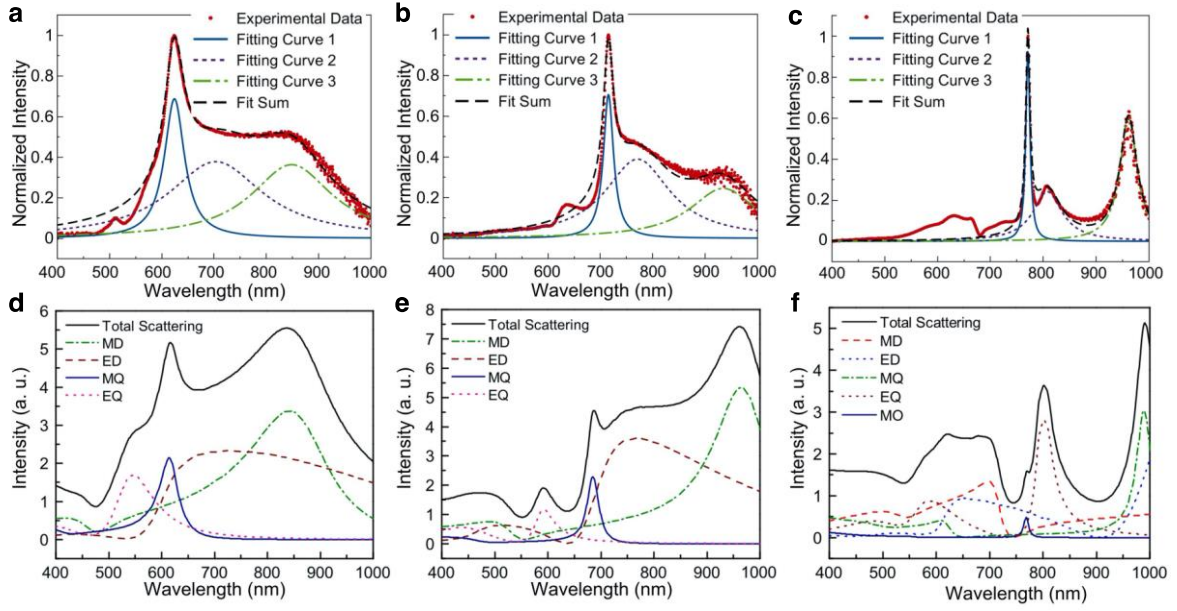

**Fig. 2 | Single-nanoparticle optical spectroscopy. a**-**c**, Scattering spectra of a single a-Si:H NP(40) (radius of ~185 nm) (**a**), a-Si:H NP(20) (radius of ~160 nm) (**b**) and a-Si:H(10) NP (radius of ~207 nm) (**c**). The red dotted curves are experimental data. Blue curves, purple dashed curves, green doted-dashed curves, and black dashed curves are fitting curves 1, 2, 3, and fit summary, respectively. **d**-**f**, Scattering spectra along with multipole decomposition calculated with Mie theory.

Mie scattering theory is used to confirm the origin of each scattering peak and clarify how the Mie resonance peaks change as a function of the hydrogen concentration. To model the permittivity of a-Si:H NPs, we use the Maxwell Garnett mixing formula

$$\varepsilon_{a\text{-Si:H}} = \varepsilon_{a\text{-Si}} \frac{(\varepsilon_{\text{voids}} + 2\varepsilon_{a\text{-Si}}) + 2f(\varepsilon_{\text{voids}} - \varepsilon_{a\text{-Si}})}{(\varepsilon_{\text{voids}} + 2\varepsilon_{a\text{-Si}}) - f(\varepsilon_{\text{voids}} - \varepsilon_{a\text{-Si}})} \quad (1)$$

where $f$ is the volume fraction of hydrogenated voids and $\varepsilon_{\text{voids}}$ denotes the permittivity of voids, which is assumed to be 1. We also slightly blue-shift the permittivity of a-Si (shown in SI) to model the hydrogenation. By using $x$=0.4 and a blue-shift of 140 nm, and $x$=0.65 and a blue-shift of 90 nm, we obtain the best matching of simulated scattering peaks with experimental data of a-Si:H NP(40) (radius of ~185 nm), a-Si:H NP(20) (radius of ~160 nm) and a-Si:H NP(10) (radius of ~207 nm), respectively.

The optical response of fabricated spherical a-Si:H NPs with dielectric permittivity $\varepsilon_{a\text{-Si:H}}$ and radius $R$ are calculated by the Mie light scattering theory[55,56], which gives the following expression for normalized scattering cross-section (SCS) for particles made of a non-magnetic material:

$$Q_{sct} = \frac{2}{(kR)^2} \sum_{l=1}^{\infty} (2l+1)\left(|a_l|^2 + |b_l|^2\right), \tag{2}$$

where $l$ defines the order of resonant mode and $k$ is the wavenumber in the surrounding material. For a single component particle, the electric ($a_l$) and magnetic ($b_l$) scattering amplitudes are given by $a_l = R_l^{(a)} / \left(R_l^{(a)} + iT_l^{(a)}\right)$, $b_l = R_l^{(b)} / \left(R_l^{(b)} + iT_l^{(b)}\right)$ and functions $R_l$ and $T_l$ can be expressed through the Bessel and Neumann functions (see Methods for details). The simulated scattering spectra are shown in Figs. 2d-f. The E-field distributions at the scattering peaks at 618 nm in Fig. 2d, 690 nm in Fig. 2e and 991 nm in Fig. 2f are shown in Supplementary Fig. 8a, b, and d, respectively. The E-field distribution profiles reveal the MQ features at these peaks, while the E-field distribution profile at 769 nm shown in Supplementary Fig. 8c displays a MO feature.

We calculated the Q factor of the measured MQ, MO and MD scattering peaks by the definition[30] $Q = \omega_0 / \Delta\omega$, discussed above. The calculated Q factor of MQ modes in Figs. 2a and b are 11 and 30, respectively. These Q factors are several times larger than those of MD modes (5 and 6 for Figs. 2a and b, respectively), suggesting that MQ modes can further boost light-matter interactions because of reduced radiation loss in a-Si:H NPs. The measured MQ scattering peaks are even more pronounced than in simulations. This implies that the actual dissipative loss of a-Si:H NPs is smaller than the one obtained by blue-shifting the $\varepsilon$ dispersion of pure a-Si. Furthermore, in Fig. 2f we observe the MO resonance with Q factor of ~100, which is ~10-20 times those obtained with plasmonic NPs at lower-order resonances (see Supplementary information section 5 for a detailed discussion).

**Highly Tunable All-Dielectric Single-Particle Nanoantenna**

To demonstrate the superiority of our results for low-loss functional nanophotonics, we show that the coupling of these high-Q scattering resonances in a single a-Si:H nanoantenna with spiropyran (SP) molecules facilitates actively tunable nanostructures possessing low tuning intensities.

In practice high input powers are needed to achieve strong all-optical tunability in Si due to its weak optical tunability[57,58]. However, the interaction length in SiNPs is limited by their nanoscale size. Thus, to achieve sufficient optical tuning of SiNPs, the high incident light intensity is needed. For instance, the intensities used for optically tuning SiNPs via electron-hole plasma excitation and laser reshaping are ~10 GW/cm$^2$ and ~21 MW/cm$^2$, respectively[59–61]. High incident light intensities lead to considerable energy consumption and can cause damage to SiNPs and surrounding materials. Therefore, all-optical tuning techniques relying on low intensities of incident light are desirable.

Here we demonstrate that coupling higher-order optical resonant modes of a-Si:H NPs with photoswitchable optical resonances of photochromic molecules[19,62,63], such as spiropyran (SP) molecules is a promising way to achieve low-intensity all-optical tuning in the visible range. The photochromism of SP is schematically displayed in Supplementary Fig. 11. The absorption spectra of molecules in SP form (blue curve) and Merocyanine (MC) form (pink curve) states are shown in Fig. 3a. Fig. 3b schematically illustrates the investigating structure composed of the tailored a-Si:H NP supporting high-Q MQ mode covered by SP molecules. The photoisomerization from SP to MC (from MC to SP) induced by UV (green) light excitation causes strong tuning of the scattering response of the single a-Si:H nanoantenna from strong scattering to suppressed scattering if the resonance of molecules in MC form is aligned with the MQ mode of the NP.

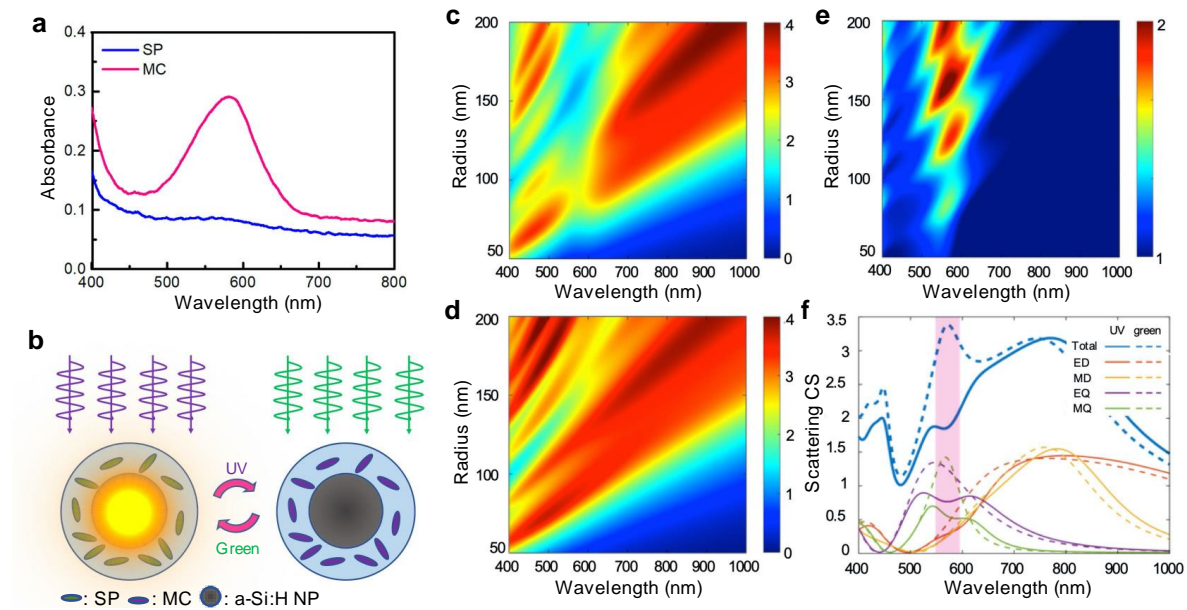

**Fig. 3 | Designing of the highly tunable photochromic all-dielectric single-particle nanoantenna. a**, Absorption spectra of molecules in SP (blue curve) and MC (pink curve) state. **b,** Schematic representation of the nanoantenna consisting of a-Si:H core and a shell of SP photochromic molecules. The nanoantenna is designed to support a higher-order Mie resonance with high Q factor at the absorption resonance of the MC state. The

photoisomerization from SP to MC state and from MC to SP state is caused by UV and green light excitation, respectively. **c,d**, Scattering cross-section (normalized to geometric) map of a core-shell nanostructure (**c**) after UV-modification and (**d**) green light recovery as a function of wavelength and a-Si:H nanoparticle radius. (**e**) Ratio between scattering cross-sections of the core-shell after green light recovery and UV-modification. (**f**) Scattering cross-section of the core-shell with the Si core radius of 130 nm and its multipole decomposition. Solid lines correspond to the core-shell after UV-modification, dashed lines correspond to the core-shell after green light recovery. The thickness of the PMMA+SP shell is 100 nm.

We start with the analytical optimization of geometry to achieve a pronounced effect of scattering tuning. To find the optimal layout, we fix the thickness of the shell (PMMA+SP) to 100 nm and vary the core radius R from 50 nm to 200 nm. For analysis, we use Mie theory for multilayered spheres[64]. The dielectric permittivity of the core is assumed to a-Si:H (40) find above, while the permittivity of PMMA+SP is obtained from experiments on transmission/reflection from the bare PMMA+SP film. The results of analytical calculations of scattering cross-section (SCS) versus R and wavelength in the MC state are summarized in Fig. 3c. We observe a pronounced dip at 580 nm in the scattering spectrum caused by the interaction with the absorption peak of MC molecules.

As the form of the photochromic molecules changes, the scattering spectrum of the nanoantenna is drastically transformed. In the fully SP form, the nanoantenna possesses scattering peaks at 580 nm, Fig. 3d, caused by different resonant modes of the a-Si:H NP(40) at different radii. The ratio of SCS in both states presented in Fig. 3e gives the change of SCS. As expected, the coupling between MQ resonance and MC molecules demonstrates a more dramatic change than for the MD resonance. We see that a large tuning effect can be achieved when the MQ resonant mode of a nanoantenna with a-Si:H NP(40) core radius of ~130 nm is at 580 nm. This drastic SCS variation is caused by the combination of a high Q factor of the MQ mode and the large dipole strength of the photochromic molecules in the SP state. Results of the mode decomposition of the structure in both states are presented in Fig 3f.

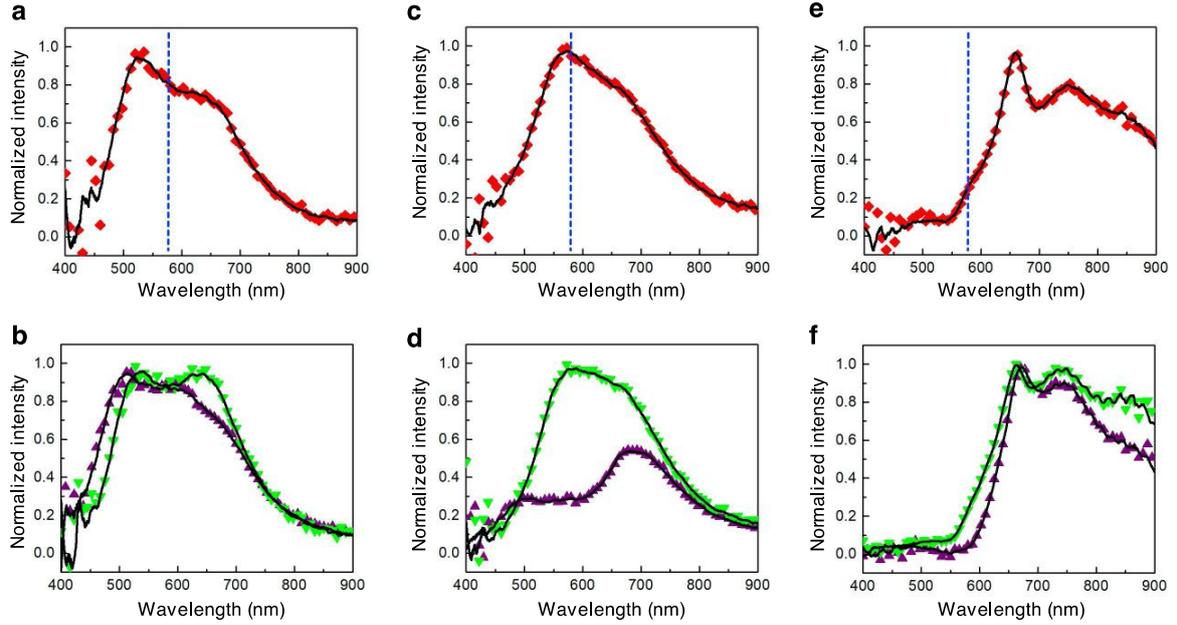

**Fig. 4 | Tuning of the all-dielectric nanoantenna**. **a**-**f**, Scattering spectra of two single a-Si:H NPs (40) (**a-d**) and a single a-Si:H NP(20) (**e, f**) covered by PMMA+SP layer. **a,c,e**, Right after the sample preparation. **b,d,f**, After the exposure to UV light (purple triangle) and after the exposure to the green light (green triangle). The vertical dashed blue dashed line illustrates the absorption peak of MC molecules.

To verify this enhanced tunability, the fabricated a-Si:H NPs have been covered by a spin-coated PMMA+SP film. Fig. 4 shows the scattering spectra of two single a-Si:H NP(40)s and one single a-Si:H NP(20) covered by PMMA+SP film. All scattering spectra have been normalized by dividing the transmission spectra of PMMA+SP film without a-Si:H NPs under the same power of dark-field light source to eliminate the contribution of bare molecular absorption. Before light irradiation, these three NPs in PMMA+SP film have MQ peaks centering at 522 nm, 573 nm, and 660 nm, respectively. The MQ peaks of the middle one match with the molecular absorption at 580 nm, while the other two have a large detuning from it. A fluorescence illuminator (mercury lamp) with bandpass filters was used to generate UV and green light, in order to switch the molecules from the SP state to the MC state and vice-versa. The estimated maximum light intensities employed in the experiment are 3.8 W/cm$^2$ and $1.1 \times 10^2$ W/cm$^2$ for UV and green light, respectively (detailed information in Methods). These intensities are significantly lower than those used in reported studies of all-optical tuning of SiNPs. For instance, the intensities used for optically tuning SiNPs via electron-hole plasma excitation and laser reshaping are ~10 GW/cm$^2$ and ~21 MW/cm$^2$, respectively[59,60].

We observe a remarkable tuning (~70%) of scattered power in the matched scenario near the MQ (Fig. 4c) scattering peak, Fig. 4d). The observed tunability is reversible and is caused by the high-Q response of the MQ resonant mode. However, when there is a detuning (Figs. 4a,e ), negligible scattering change occurs when SP molecules are switched to MC state, i.e., after UV light exposure (Fig. 4b,f). This change vanishes when the molecules are switched back to SP form by irradiating the sample with green light, as shown by Fig. 4b, d, f. This photoswitchable variation confirms that the strong modulation of MQ scattering observed here is derived from the interaction between MQ mode and photochromic molecules.

## Conclusions

In summary, we have used bandgap engineering to tailor a-Si:H NPs with low dissipation losses in the visible range. As a result, strong higher-order magnetic multipole resonances in the visible and NIR ranges with Q factors up to 100 have been experimentally demonstrated in tailored a-Si:H NPs. The appearance of strong magnetic multipole scattering modes stems from the larger bandgap of a-Si:H NPs compared to pure SiNPs. The bandgap of a-Si:H NPs can be tuned by changing the hydrogen concentration. Strong magnetic multipole scattering modes with narrow linewidths in the visible and NIR regions open new opportunities to use nanostructures made of low-loss high-index materials for light manipulation, exploring light-matter interactions and tunable optical devices. Remarkably, we show that coupling such high-Q scattering resonances in a single a-Si:H nanoantenna with photochromic SP molecules results in ~70% tuning of the scattering peak intensity upon switching the photochromic molecules between transparent SP and colored MC states. This photochromatic all-optical tuning of SiNPs requires drastically lower incident light intensity than other reported methods. Our results pave the way to novel devices based on highly tunable all-dielectric single nanostructures with low-intensity requirements.

## Methods

**Sample preparation**

A 10 ml titanium batch reactor (High-Pressure Equipment Company (HiP Co.)) was used for the synthesis. First, 21 µL trisilane ($Si_3H_8$, 100%, Voltaix) and n-hexane (anhydrous, 95%, Sigma-Aldrich) were loaded in the reactor in a nitrogen-filled glovebox. The amount of n-hexane loaded in the reactor is associated with the reaction pressure inside the reactor during the heating process. In all reactions, the pressure was kept at 34.5 MPa (5000 psi). The different hydrogen concentration in these a-Si:H NPs is determined by different reaction temperatures[52]. For example, the a-Si:H nanoparticles with a hydrogen concentration of 40% were synthesized at a temperature of 380 °C[52]. After adding the reagents, the reactor was sealed by using a wrench inside the glove box. Then a vice was used to tightly seal the reactor after removing it from the glove box. The reactor was heated to the target temperature in a heating block for 10 min to allow the complete decomposition of trisilane. After the reaction, an ice bath was used to cool the reactor to room temperature. Colloidal a-Si:H NPs were then extracted from the opened reactor. The nanoparticles were washed by chloroform (99.9%, Sigma-Aldrich) using a centrifuge (at 8000 rpm for 5 min).

The SP+PMMA film is prepared by mixing spiropyran (Sigma-Aldrich) molecules with PMMA (Sigma-Aldrich) with a weight ratio of 1:1 in chlorobenzene (2 wt % of SP in chlorobenzene). Then a spin coater (Laurell) is used to coat the mixture on a-Si:H NPs at 2000 rpm for 1 min.

**Optical measurements**

Before the scattering measurement, a-Si:H NPs were drop coasted on a bare glass substrate. An inverted microscope (Ti-E, Nikon) with a spectrograph (Andor), an EMCCD (Andor) and a halogen white light source (12V, 100 W) was employed to measure the scattering spectra of single a-Si:H NPs on the bare glass substrate[18,19]. Raman spectra of a-Si:H NPs were measured by the Witec Micro-Raman Spectrometer. A UV-VIS-NIR spectrometer (Ocean Optics) was used to measure the transmission of a-Si:H NPs in ethanol. An Epi-fluorescence illuminator (mercury lamp, Nikon) and bandpass filters (central wavelength: 350 nm (UV) and 540 nm (green), bandwidth: 50 nm (UV) or 25 nm (green)) are used to generated UV and green light. The maximum power of UV and green light was measured by a power meter (Thorlabs). Since the mercury lamp is an incoherent light source and the incident light covers the whole rear aperture of the objective, the beam spot diameter used to calculate the incident light intensity is obtained by: $d_{spot}=d_{aperture}$/magnification, where $d_{aperture}$ is the diameter of the objective rear aperture and =6.5 mm and magnification is 100 times.

**Analytical and numerical simulations**

The optical response of a Si nanoparticle with dielectric permittivity $\varepsilon = n^2$ ($n$ is the refractive index of the nanoparticle material) and a radius $R$ located in the free space can be treated via Mie light scattering theory[55,56], which gives the following expression for normalized scattering [$Q_{sct} = P_{sct}/(\pi R^2 I)$] cross-section for particles made of a non-magnetic material:

$$Q_{sct} = \frac{2}{(kR)^2} \sum_{l=1}^{\infty} (2l+1)\left(|a_l|^2 + |b_l|^2\right), \qquad (3)$$

where $l$ defines the order of partial wave, $k$ is the wavenumber $k = \omega n_h / c$, $\lambda$ is the wavelength in a vacuum and $\varepsilon_h = n_h^2$ is the dielectric permittivity of the surrounding medium. The quantity $P_{sct}$ denotes the scattering power, $I$ is the excitation intensity, $c$ is the speed of light, and $\varepsilon_0$ is the dielectric constant.

For a single component particle, the electric and magnetic scattering amplitudes are given by

$$a_l = \frac{R_l^{(a)}}{R_l^{(a)} + i T_l^{(a)}}, \quad b_l = \frac{R_l^{(b)}}{R_l^{(b)} + i T_l^{(b)}}, \qquad (4)$$

and functions $R_l$ and $T_l$ can be expressed in the following form:

$$R_l^{(a)} = n\psi_l'(kR)\psi_l(nkR) - \psi_l(kR)\psi_l'(nkR),$$
$$T_l^{(a)} = n\chi_l'(kR)\psi_l(nkR) - \chi_l(kR)\psi_l'(nkR), \qquad (5)$$

$$R_l^{(b)} = n\psi_l'(nkR)\psi_l(kR) - \psi_l(nkR)\psi_l'(kR),$$
$$T_l^{(b)} = n\chi_l(kR)\psi_l'(nkR) - \chi_l'(kR)\psi_l(nkR). \qquad (6)$$

Here, $\psi_l(x) = \sqrt{\frac{\pi x}{2}} J_{l+1/2}(x)$, $\chi_l(x) = \sqrt{\frac{\pi x}{2}} N_{l+1/2}(x)$, $J_{l+1/2}(x)$ and $N_{l+1/2}(x)$ are the Bessel and Neumann functions, and prime means derivation.

The Mie scattering coefficients can be expressed similarly and can be found in Refs.[55,65].

The optical properties of the nanostructures in the optical frequency range have been studied numerically by using CST Microwave Studio. CST Microwave Studio is a full-wave 3D electromagnetic field solver based on a finite-integral time-domain solution technique. A nonuniform mesh was used to improve accuracy in the vicinity of the Si nanoparticles where the field concentration was significantly large.

# References

1. Ma, R. M. & Oulton, R. F. Applications of nanolasers. *Nat. Nanotechnol.* **14**, 12–22 (2019).

2. Krasnok, A., Caldarola, M., Bonod, N. & Alú, A. Spectroscopy and Biosensing with Optically Resonant Dielectric Nanostructures. *Adv. Opt. Mater.* **6**, 1701094 (2018).

3. Krasnok, A., Tymchenko, M. & Alù, A. Nonlinear metasurfaces: a paradigm shift in nonlinear optics. *Mater. Today* **21**, 8–21 (2018).

4. Armani, D. K., Kippenberg, T. J., Spillane, S. M. & Vahala, K. J. Ultra-high-Q toroid microcavity on a chip. *Nature* **421**, 925–928 (2003).

5. Zhang, N. *et al.* High-: Q and highly reproducible microdisks and microlasers. *Nanoscale* **10**, 2045–2051 (2018).

6. Kryzhanovskaya, N. *et al.* Enhanced light outcoupling in microdisk lasers via Si spherical nanoantennas. *J. Appl. Phys.* **124**, 163102 (2018).

7. Moiseev, E. I. *et al.* Light Outcoupling from Quantum Dot-Based Microdisk Laser via Plasmonic Nanoantenna. *ACS Photonics* **4**, 275–281 (2017).

8. McCall, S. L., Levi, A. F. J., Slusher, R. E., Pearton, S. J. & Logan, R. A. Whispering-gallery mode microdisk lasers. *Appl. Phys. Lett.* **60**, 289–291 (1992).

9. Vassiliev, V. V. *et al.* Narrow-line-width diode laser with a high-Q microsphere resonator. *Opt. Commun.* **158**, 305–312 (1998).

10. Koschorreck, M. *et al.* Dynamics of a high-Q vertical-cavity organic laser. *Appl. Phys. Lett.* **87**, 181108 (2005).

11. Lončar, M., Yoshie, T., Scherer, A., Gogna, P. & Qiu, Y. Low-threshold photonic crystal laser. *Appl. Phys. Lett.* **81**, 2680–2682 (2002).

12. Song, B.-S., Noda, S., Asano, T. & Akahane, Y. Ultra-high-Q photonic double-heterostructure nanocavity. *Nat. Mater.* **4**, 207–210 (2005).

13. Painter, O. *et al.* Two-dimensional photonic band-gap defect mode laser. *Science (80-. ).* **284**, 1819–1821 (1999).

14. Atwater, H. A. & Polman, A. Plasmonics for improved photovoltaic devices. *Nat. Mater.* **9**, 205–213 (2010).

15. Schuller, J. A. *et al.* Plasmonics for extreme light concentration and manipulation. *Nat. Mater.* **9**, 368–368 (2010).

16. Zia, R., Schuller, J. A., Chandran, A. & Brongersma, M. L. Plasmonics: the next chip-scale technology. *Mater. Today* **9**, 20–27 (2006).

17. Brongersma, M. L., Halas, N. J. & Nordlander, P. Plasmon-induced hot carrier science and technology. *Nat. Nanotechnol.* **10**, 25–34 (2015).

18. Wang, M. *et al.* Molecular-Fluorescence Enhancement via Blue-Shifted Plasmon-Induced Resonance Energy Transfer. *J. Phys. Chem. C* **120**, 14820–14827 (2016).

19. Wang, M. *et al.* Controlling Plasmon-Enhanced Fluorescence via Intersystem Crossing in Photoswitchable Molecules. *Small* **13**, 1701763 (2017).

20. Wang, M. *et al.* Plasmon-trion and plasmon-exciton resonance energy transfer from a single plasmonic nanoparticle to monolayer MoS2. *Nanoscale* **9**, 13947–13955 (2017).

21. Fan, X., Zheng, W. & Singh, D. J. Light scattering and surface plasmons on small spherical particles. *Light Sci. Appl.* **3**, e179–e179 (2014).

22. Hsu, C. W., Zhen, B., Stone, A. D., Joannopoulos, J. D. & Soljačić, M. Bound states in the continuum. *Nat. Rev. Mater.* **1**, 16048 (2016).

23. Rybin, M. V. *et al.* High- Q Supercavity Modes in Subwavelength Dielectric

Resonators. *Phys. Rev. Lett.* **119**, 243901 (2017).

24. Krasnok, A. *et al.* Anomalies in light scattering. *Adv. Opt. Photonics* **11**, 892 (2019).

25. Monticone, F., Sounas, D. L., Krasnok, A. & Alù, A. Can a nonradiating mode be externally excited? Nonscattering states vs. embedded eigenstates. *ACS Photonics* acsphotonics.9b01104 (2019). doi:10.1021/acsphotonics.9b01104

26. Wang, M. *et al.* Plasmon–trion and plasmon–exciton resonance energy transfer from a single plasmonic nanoparticle to monolayer $MoS_2$. *Nanoscale* **9**, 13947–13955 (2017).

27. Maier, S. A. & Atwater, H. A. Plasmonics: Localization and guiding of electromagnetic energy in metal/dielectric structures. *J. Appl. Phys.* **98**, 011101 (2005).

28. Wang, H., Brandl, D. W., Nordlander, P. & Halas, N. J. Plasmonic Nanostructures: Artificial Molecules. *Acc. Chem. Res.* **40**, 53–62 (2007).

29. Halas, N. J., Lal, S., Chang, W. S., Link, S. & Nordlander, P. Plasmons in strongly coupled metallic nanostructures. *Chem. Rev.* **111**, 3913–3961 (2011).

30. Baranov, D. G. *et al.* All-dielectric nanophotonics: the quest for better materials and fabrication techniques. *Optica* **4**, 814 (2017).

31. Genevet, P., Capasso, F., Aieta, F., Khorasaninejad, M. & Devlin, R. Recent advances in planar optics: from plasmonic to dielectric metasurfaces. *Optica* **4**, 139 (2017).

32. Jahani, S. & Jacob, Z. All-dielectric metamaterials. *Nat. Nanotechnol.* **11**, 23–36 (2016).

33. Krasnok, A. E., Miroshnichenko, A. E., Belov, P. A. & Kivshar, Y. S. All-dielectric optical nanoantennas. *Opt. Express* **20**, 20599 (2012).

34. Krasnok, A. E., Filonov, D. S., Simovski, C. R., Kivshar, Y. S. & Belov, P. A. Experimental demonstration of superdirective dielectric antenna. *Appl. Phys. Lett.* **104**, (2014).

35. Kuznetsov, A. I., Miroshnichenko, A. E., Brongersma, M. L., Kivshar, Y. S. & Luk'yanchuk, B. Optically resonant dielectric nanostructures. *Science (80-. ).* **354**, (2016).

36. Ha, S. T. *et al.* Directional lasing in resonant semiconductor nanoantenna arrays. *Nat. Nanotechnol.* **13**, 1042–1047 (2018).

37. Staude, I. & Schilling, J. Metamaterial-inspired silicon nanophotonics. *Nat. Photonics* **11**, 274–284 (2017).

38. Makarov, S. *et al.* Tuning of Magnetic Optical Response in a Dielectric Nanoparticle by Ultrafast Photoexcitation of Dense Electron-Hole Plasma. *Nano Lett.* **15**, 6187–6192 (2015).

39. Baranov, D. G., Makarov, S. V., Krasnok, A. E., Belov, P. A. & Alù, A. Tuning of near- and far-field properties of all-dielectric dimer nanoantennas via ultrafast electron-hole plasma photoexcitation. *Laser Photonics Rev.* **10**, 1009–1015 (2016).

40. Bohn, J. *et al.* Active Tuning of Spontaneous Emission by Mie-Resonant Dielectric Metasurfaces. *Nano Lett.* **18**, 3461–3465 (2018).

41. Shcherbakov, M. R. *et al.* Ultrafast All-Optical Switching with Magnetic Resonances

in Nonlinear Dielectric Nanostructures. *Nano Lett.* **15**, 6985–6990 (2015).

42. Rahmani, M. *et al.* Reversible Thermal Tuning of All-Dielectric Metasurfaces. *Adv. Funct. Mater.* **27**, 1–7 (2017).

43. Jang, J. *et al.* Kerker-Conditioned Dynamic Cryptographic Nanoprints. *Adv. Opt. Mater.* 1801070 (2018). doi:10.1002/adom.201801070

44. Fu, Y. H., Kuznetsov, A. I., Miroshnichenko, A. E., Yu, Y. F. & Luk'yanchuk, B. Directional visible light scattering by silicon nanoparticles. *Nat. Commun.* **4**, 1–6 (2013).

45. Kuznetsov, A. I., Miroshnichenko, A. E., Fu, Y. H., Zhang, J. & Luk'yanchuk, B. Magnetic light. *Sci. Rep.* **2**, 492 (2012).

46. Evlyukhin, A. B. *et al.* Demonstration of magnetic dipole resonances of dielectric nanospheres in the visible region. *Nano Lett.* **12**, 3749–3755 (2012).

47. Tribelsky, M. I. & Luk'yanchuk, B. S. Anomalous Light Scattering by Small Particles. *Phys. Rev. Lett.* **97**, 263902 (2006).

48. Bücher, K., Bruns, J. & Wagemann, H. G. Absorption coefficient of silicon: An assessment of measurements and the simulation of temperature variation. *J. Appl. Phys.* **75**, 1127–1132 (1994).

49. Ning, C.-Z., Dou, L. & Yang, P. Bandgap engineering in semiconductor alloy nanomaterials with widely tunable compositions. *Nat. Rev. Mater.* **2**, 17070 (2017).

50. Khurgin, J. B. & Sun, G. In search of the elusive lossless metal. *Appl. Phys. Lett.* **96**, 181102 (2010).

51. Dong, W. *et al.* Wide Bandgap Phase Change Material Tuned Visible Photonics. *Adv. Funct. Mater.* **29**, 1806181 (2019).

52. Harris, J. T., Hueso, J. L. & Korgel, B. A. Hydrogenated amorphous silicon (a-Si:H) colloids. *Chem. Mater.* **22**, 6378–6383 (2010).

53. Smets, A. H. M. *et al.* The Relation Between the Bandgap and the Anisotropic Nature of Hydrogenated Amorphous Silicon. *IEEE J. Photovoltaics* **2**, 94–98 (2012).

54. Pell, L. E., Schricker, A. D., Mikulec, F. V. & Korgel, B. A. Synthesis of amorphous silicon colloids by trisilane thermolysis in high temperature supercritical solvents. *Langmuir* **20**, 6546–6548 (2004).

55. Savelev, R. S., Sergaeva, O. N., Baranov, D. G., Krasnok, A. E. & Alù, A. Dynamically reconfigurable metal-semiconductor Yagi-Uda nanoantenna. *Phys. Rev. B* **95**, 1–9 (2017).

56. Bohren, Craig F, Huffman, D. R. *Absorption and Scattering of Light by Small Particles*. *Wiley Science* (Wiley-VCH Verlag GmbH, 1998). doi:10.1002/9783527618156

57. Leuthold, J., Koos, C. & Freude, W. Nonlinear silicon photonics. *Nat. Photonics* **4**, 535–544 (2010).

58. Pala, R. A., Shimizu, K. T., Melosh, N. A. & Brongersma, M. L. A Nonvolatile Plasmonic Switch Employing Photochromic Molecules. *Nano Lett.* **8**, 1506–1510

(2008).

59. Zograf, G. P. *et al.* Resonant Nonplasmonic Nanoparticles for Efficient Temperature-Feedback Optical Heating. *Nano Lett.* **17**, 2945–2952 (2017).

60. Baranov, D. G. *et al.* Nonlinear Transient Dynamics of Photoexcited Resonant Silicon Nanostructures. *ACS Photonics* **3**, 1546–1551 (2016).

61. Zuev, D. A. *et al.* Fabrication of Hybrid Nanostructures via Nanoscale Laser-Induced Reshaping for Advanced Light Manipulation. *Adv. Mater.* **28**, 3087–3093 (2016).

62. Klajn, R. Spiropyran-based dynamic materials. *Chem. Soc. Rev.* **43**, 148–184 (2014).

63. Lin, L. *et al.* Photoswitchable Rabi Splitting in Hybrid Plasmon-Waveguide Modes. *Nano Lett.* **16**, 7655–7663 (2016).

64. Alù, A. & Engheta, N. Polarizabilities and effective parameters for collections of spherical nanoparticles formed by pairs of concentric double-negative, single-negative, and∕or double-positive metamaterial layers. *J. Appl. Phys.* **97**, 094310 (2005).

65. Aden, A. L. & Kerker, M. Scattering of Electromagnetic Waves from Two Concentric Spheres. *J. Appl. Phys.* **22**, 1242–1246 (1951).
## Acknowledgements

Y.Z. and J.F. acknowledge the financial support of the National Aeronautics and Space Administration Early Career Faculty Award (80NSSC17K0520), and the National Institute of General Medical Sciences of the National Institutes of Health (DP2GM128446). M.S. acknowledges the financial support of University Graduate Continuing Fellowship of the University of Texas at Austin. M.S., A.K. and A.A. acknowledge the financial support of the Air Force Office of Scientific Research and the National Science Foundation.
## Author information

**Affiliations**

Department of Mechanical Engineering and Texas Materials Institute, The University of Texas at Austin, Austin, Texas, United States

M. Wang, J. Fang & Y. Zheng

Photonics Initiative, Advanced Science Research Center, City University of New York, New York, USA

A. Krasnok, G. Hu & A. Alù
17

ITMO University, St. Petersburg 197101, Russia

S. Lepeshov

McKetta Department of Chemical Engineering, The University of Texas at Austin, Austin, Texas, United States.

T. Jiang & B. A. Korgel

Department of Electrical and Computer Engineering, National University of Singapore, Singapore, Singapore.

G. Hu

**Contributions**

Y.Z, A.A., M.W., and A.K. conceived and coordinated the projects. M.W. performed experiments with the assistance of J.F.. A.K., S.L. and G.H. performed the analytical and numerical simulations. T. J. and B.A.K. synthesized the a-Si:H NPs. M.W. and A.K. wrote the manuscript. All authors discussed the results and commented on the manuscript.

**Corresponding author**

Correspondence to Y. Zheng, A. Alù and A. Krasnok.

# Supplementary information

**Supplementary Information**

Supplementary information sections 1-5, Figs. 1–11 and refs 1-4.

# Supplementary Information:

# Suppressing Material Loss for Functional Nanophotonics Using Bandgap Engineering

Mingsong Wang[1,2], Alex Krasnok[2,*], Sergey Lepeshov[3], Guangwei Hu[2,5], Taizhi Jiang[4], Jie Fang[1], Brian A. Korgel[4], Andrea Alù[2,*], and Yuebing Zheng[1*]

[1]*Walker Department of Mechanical Engineering and Texas Materials Institute, The University of Texas at Austin, Austin, Texas 78712, United States*
[2]*Photonics Initiative, Advanced Science Research Center, City University of New York, New York 10031, USA*
[3]*ITMO University, St. Petersburg 197101, Russia*
[4]*McKetta Department of Chemical Engineering, The University of Texas at Austin, Austin, Texas 78712, United States.*
[5]*Department of Electrical and Computer Engineering, National University of Singapore, Singapore 117583, Singapore.*
akrasnok@gc.cuny.edu, aalu@gc.cuny.edu, zheng@austin.utexas.edu

## 1. SEM images of a-Si:H NPs

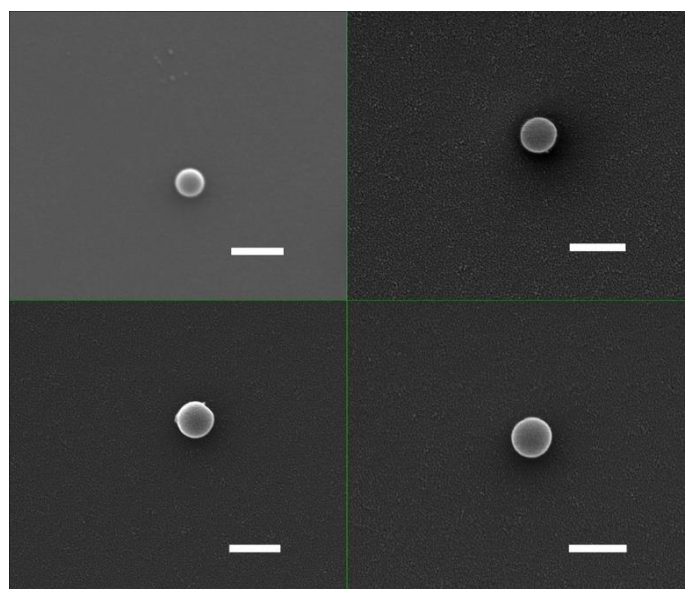

**Supplementary Fig. S1**│ SEM images of a-Si:H NPs with 40% hydrogen (a-Si:H(40)). Scale bar is 500 nm.

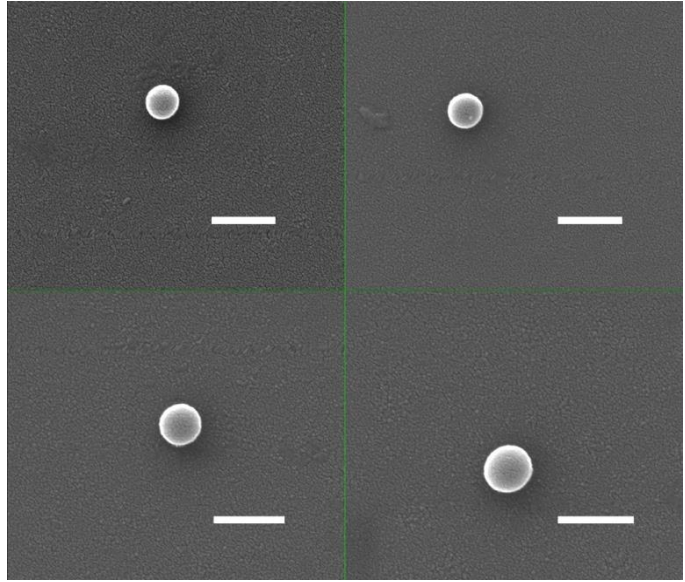

**Supplementary Fig. S2** SEM images of a-Si:H NPs with 20% hydrogen (a-Si:H(20)). Scale bar is 500 nm.

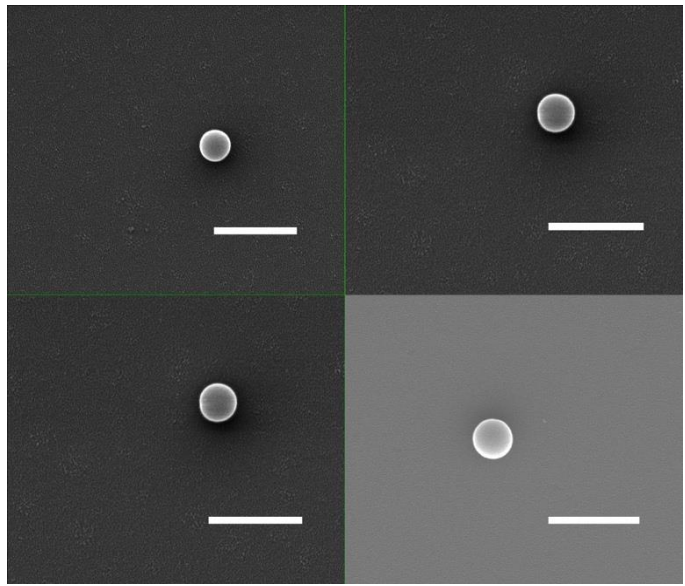

**Supplementary Fig. S3** SEM images of a-Si:H NPs with 10% hydrogen (a-Si:H(10)). Scale bar is 1 μm.

## 2. Transmission spectra of a-Si:H NPs

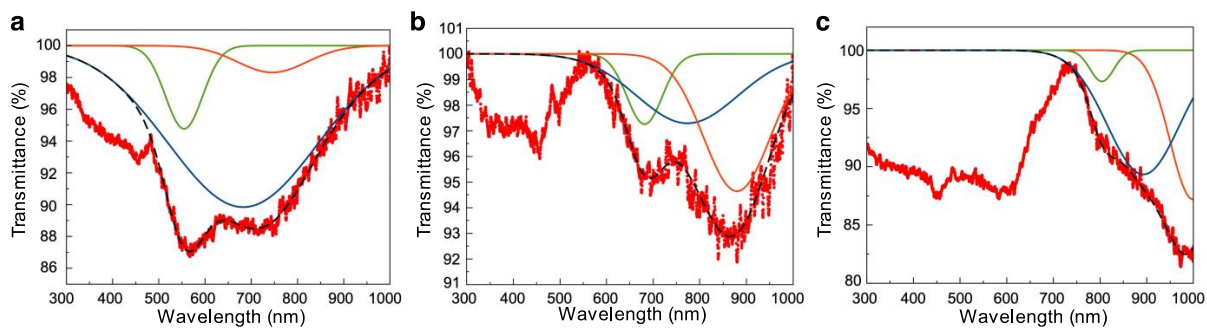

**Supplementary Fig. S4** **a**,**b**,**c**, Transmission spectra of a-Si:H NPs with 40% (a), 20% (b) and 10% (c) hydrogen. Green curves, red curves, blue curves are fitting Gaussian curves for magnetic quadrupole mode (MQ), electric dielectric dipole mode (ED), and magnetic dipole mode (MD), respectively. Black dashed curves are fit summary.

a-Si:H NPs can strongly scatter light due to the excitation of Mie resonances[1,2], so transmission spectra show the optical extinction possessing both absorption and scattering signals. Supplementary Fig. 5 shows that transmission spectra of a-Si:H NPs have several dips, those dips at longer wavelength are brought by optical scatterings and can be fitted into three Gaussian peaks. It should be mentioned that Gaussian distribution instead of Lorentzain distribution being used for fitting transmission dips is because transmission spectra were obtained from multiple particles in the solvent. Fitting dip positions in Supplementary Fig. 5 match well with corresponding magnetic quadrupole (MQ), electric dipole (ED) and magnetic dipole (MD), magnetic octupole (MO), and electric quadrupole (EQ) scattering peak positions of a-Si:H NPs (40, 20 and 10), as shown in Fig. 2. The bare optical absorption of a-Si:H NPs can be gained by removing those scattering signals.

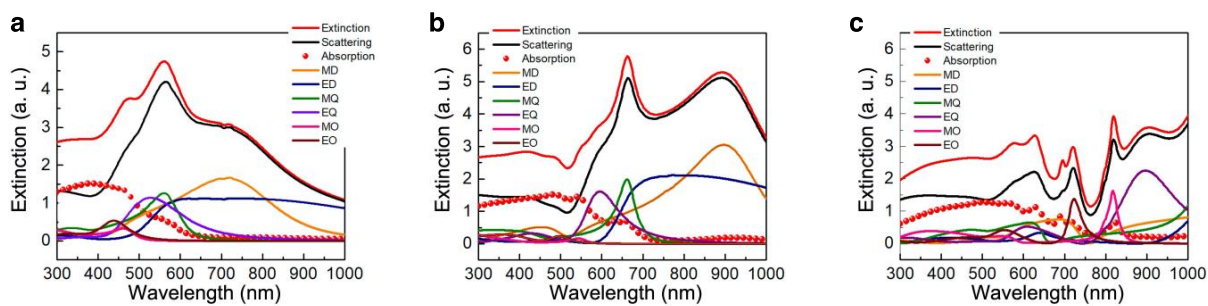

**Supplementary Fig. S5** **a**,**b**,**c**, Simulated extinction spectra of a-Si:H NPs(40) (diameters: 300 nm) (**a**), a-Si:H NPs (20) (280 nm) (**b**), and a-Si:H NPs(10) (diameters: 420 nm) (**c**). EO represents the electric octupole.

In order to further verify the optical absorption of a-Si:H NPs, we simulated the extinction spectra of a-Si:H NPs with different hydrogen concentrations through Mie theory and used the same method mentioned in the main text to obtain their permittivity and dispersion. Since the transmission spectra were obtained through ensemble measurements, we chose a-Si:H NPs (40, 20 and 10) having three representative sizes (diameters: 300nm, 280nm and 420 nm) to calculate extinction spectra, respectively. The obtained extinction spectra with absorption components and absorption components are shown in Supplementary Fig. 6. The simulated spectra are in good agreement with experimental results.

## 3. Optical setup and E-field distribution

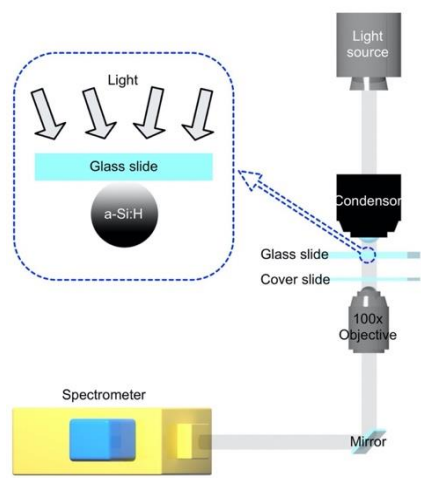

**Supplementary Fig. S6**│ Schematics of the optical setup and sample configuration for the single-nanoparticle dark-field scattering spectroscopy.

## 4. Permittivity dispersion of a-Si and E-field distribution

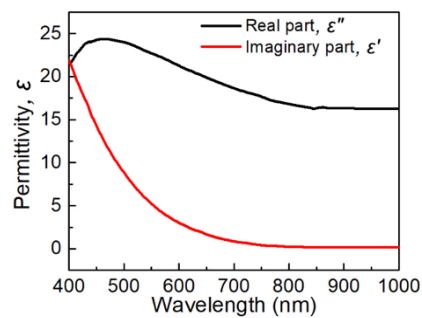

**Supplementary Fig. S7** Dependence of permittivity dispersion ($\varepsilon$) of a-Si on the wavelength. Permittivity contains real part ($\varepsilon'$) and imaginary part ($\varepsilon''$): $\varepsilon = \varepsilon' + i\varepsilon''$.

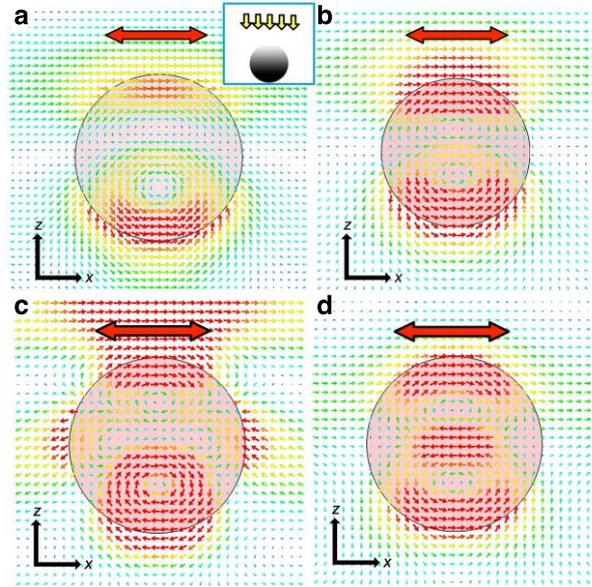

**Supplementary Fig. S8** **a,b**, Numerically calculated E-field distribution profiles at 616 nm (**a**) in Fig. 2d and at 687 nm (**b**) in Fig. 2e. **c,d**, Numerically calculated E-field distribution profiles at 769 nm (**c**) and at 991 nm (**d**) in Fig. 2f. The incident light (yellow arrows) irradiates on a-Si:H NPs from the top as shown by the inset, and the red arrow indicates the light polarization.

## 5. Demonstration of MQ and MO modes

Scattering spectra of a-Si:H (40 and 20) NPs with different diameters are shown in Supplementary Fig. 9. These results show that the MQ scattering generally diminishes when the scattering peak blueshifts towards the absorption band of a-Si:H NPs, where dissipative loss and the imaginary part of permittivity ($\varepsilon''$) increase significantly. It is also observed that the MQ scattering fades faster than the MD scattering, as expected because MQ modes have a larger stored energy, and therefore they are more prone to loss. This explains the lack of observation of these modes in previous experiments using conventional materials. In order to statistically reveal this trend, we show the dependence of the ratio between MQ and MD scattering peak intensity versus wavelength in Supplementary Figs. 9c and d. A steady decrease of MQ vs MD ratio is observed when the wavelength shifts towards the absorption peak of a-Si:H NPs.

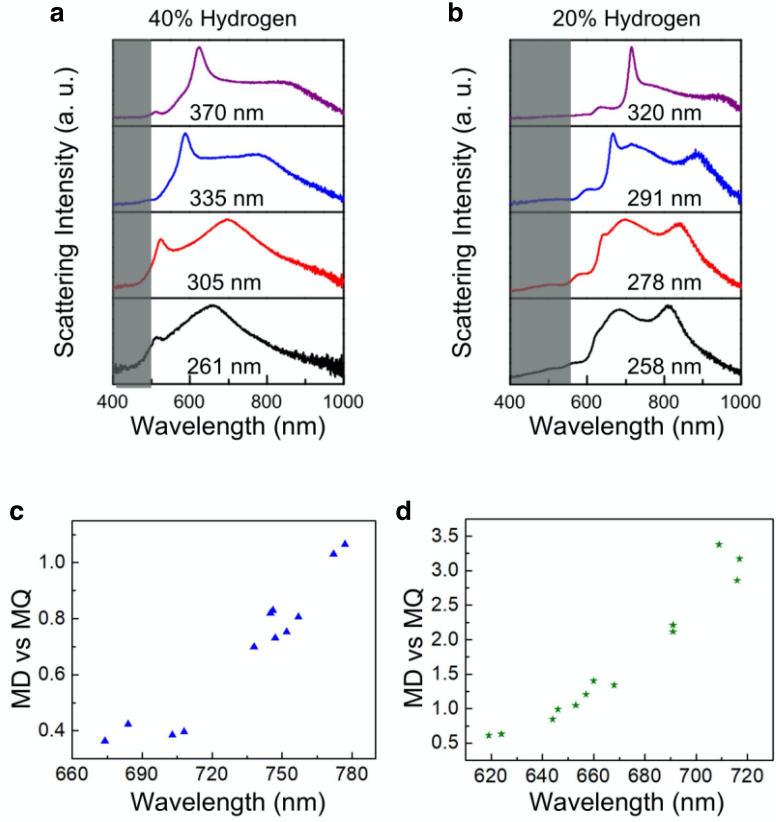

**Supplementary Fig. S9** | **a**, Scattering spectra of single a-Si:H(40) NPs. Diameters of NPs are 261 nm (black curve), 305 nm (red curve), 335 nm (blue curve) and 370 nm (purple curve). **b**, Scattering spectra of single a-Si:H(20) NPs. Diameters of NPs are 258 nm (black curve), 278 nm (red curve), 291 nm (blue curve) and 320 nm (purple curve). **c,d**, Dependences of the ratio between MQ and MD scattering peak intensity on the wavelength for a-Si:H(40) NPs (**c**) and a-Si:H(20) NPs (**d**).

We can also draw the same conclusion by comparing a-Si:H NPs with different hydrogen concentrations. Both the blue curve in Supplementary Fig. 9a and the back curve in Supplementary Fig. 9b show a MQ resonance at around 600 nm and a MD resonance at about 800 nm, while the MQ vs MD ratio of a-Si:H(40) NPs is larger than the one of a-Si:H(20), because the absorption of a-Si:H(40) NPs is at a shorter wavelength. As an additional remarkable knob, these results indicate that the minimum wavelength of the MQ scattering peak can be tuned by varying the hydrogen concentration of a-Si:H NPs.

By further comparing Supplementary Fig. 9a with Supplementary Fig. 9b, it is observed that both MQ and MD scattering peaks are narrower when the hydrogen concentration is higher. The linewidth of the scattering peaks follows[3]:

$$\gamma_\ell = \frac{q^{2\ell+1}(\ell+1)}{[\ell(2\ell-1)!!]^2 (d\epsilon/d\omega)_\ell} \qquad (S1)$$

where $q$ is a size parameter and equates $2\pi R/\lambda$, $\ell$ is the order of the optical mode, and $\omega$ is frequency. $\epsilon = \epsilon_p(\omega)/\epsilon_m(\omega)$, where $\epsilon_p(\omega)$ and $\epsilon_m(\omega)$ is the permittivity of the particle and the medium, respectively. This equation reveals that the value of $\gamma_\ell$ decreases with a decrease in $q$ and an increase in $\ell$, assuming that $\epsilon$ and $\omega$ are fixed. When $n$ is larger than $2^{1,4}$, $2R$ is approximately proportional to $\lambda/n$ at Mie resonance. Therefore, $q$ is inversely proportional to $n$ at the MQ resonances; i.e., $\gamma_\ell$ decreases when $n$ becomes larger. Since the decrease in hydrogen concentration leads to a larger refractive index, a-Si:H NPs with higher hydrogen concentration have narrower MQ scatterings and higher Q-factors.

Higher-order multipole scattering modes, such as the magnetic octupole (MO) scattering, were also observed in the NIR region by further reducing the hydrogen concentration and enlarging the particle size. Fig. 2c shows scattering spectra of an a-Si:H(10) NP with a diameter of ~414 nm, the major three peaks can be again fitted with Lorentzian models. Simulation results (particle diameter: 400 nm, $x$=0.7 and a 50 nm blueshift of $\varepsilon$ dispersion) in Fig. 2f reveal that the three peaks at 769 nm, 801 nm and 991 nm are MO, EQ and MQ, respectively. Scattering spectra of a-Si:H(10) NPs with diameter from 356 nm to 438 nm are shown in Supplementary Fig. 10a. These spectra highlight that MO scattering is stronger than MQ scattering when it is away from the absorption peak and MO gradually decreases when the peaks shift towards the absorption peak, as illustrated by the dependence of MO vs MQ ratio on the wavelength for a-Si:H(10) in Supplementary Fig. 10b. We calculated the Q factors of the MO scattering peaks in Supplementary Fig. 10a, and obtained the highest Q factor of 100, which is ~10-20 times of those obtained with plasmonic NPs at lower-order resonances.

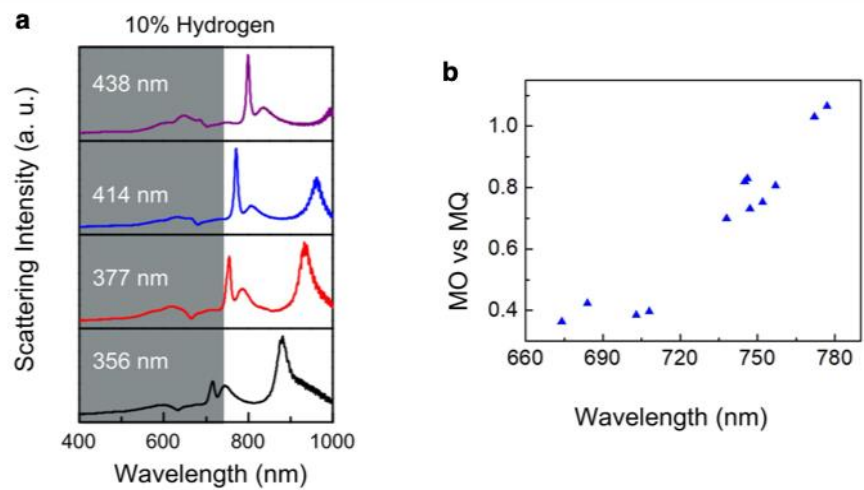

**Supplementary Fig. S10** ⎪ **a**, Scattering spectra of single a-Si:H(10) NPs. Sizes of NPs are 356 nm (black curve), 377 nm (red curve), 414 nm (blue curve) and 438 nm (purple curve). **b**, Dependence of MO vs MQ ratio on the wavelength for a-Si:H(10).

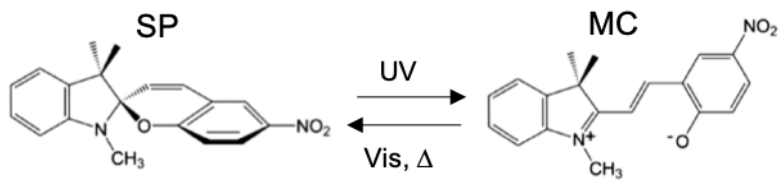

**Supplementary Fig. S11** ⎪ Photochromism of spiropyran.

# References

1. Kuznetsov, A. I., Miroshnichenko, A. E., Fu, Y. H., Zhang, J. & Luk'yanchuk, B. Magnetic light. *Sci. Rep.* **2**, 492 (2012).

2. Fu, Y. H., Kuznetsov, A. I., Miroshnichenko, A. E., Yu, Y. F. & Luk'yanchuk, B. Directional visible light scattering by silicon nanoparticles. *Nat. Commun.* **4**, 1–6 (2013).

3. Tribelsky, M. I. & Luk'yanchuk, B. S. Anomalous Light Scattering by Small Particles. *Phys. Rev. Lett.* **97**, 263902 (2006).

4. Kuznetsov, A. I., Miroshnichenko, A. E., Brongersma, M. L., Kivshar, Y. S. & Luk'yanchuk, B. Optically resonant dielectric nanostructures. *Science* **354**, aag2472 (2016).